\documentclass[twocolumn,showpacs,showkeys,preprintnumbers,amsmath,amssymb]{revtex4}

\usepackage{graphicx}
\usepackage{dcolumn}
\usepackage{bm}

\begin{document}

\setlength{\topmargin}{0pt}
\setlength{\parskip}{0.5\baselineskip}

\preprint{APS/PRB}

\title{Hidden magnetic transitions in
  thermoelectric layered cobaltite,
  [Ca$_2$CoO$_3$]$_{0.62}$[CoO$_2$]
  }

\author{J. Sugiyama$^1$}
  \email{sugiyama@iclab.tytlabs.co.jp}
\author{J. H. Brewer$^2$}
\author{E. J. Ansaldo$^3$}
\author{H. Itahara$^1$}%
\author{K. Dohmae$^1$}%
\author{Y. Seno$^1$}%
\author{C. Xia$^1$}%
\author{T. Tani$^1$}%
\affiliation{%
$^1$Toyota Central Research and Development Labs. Inc.,
  Nagakute, Aichi 480-1192, Japan}%

\affiliation{
$^2$TRIUMF, CIAR and
Department of Physics and Astronomy, University of British Columbia,
Vancouver, BC, V6T 1Z1 Canada
}%

\affiliation{
$^3$Department of Physics, University of Saskatchewan,
  Saskatoon, SK, S7N 5A5 Canada
}%

\date{\today}

\begin{abstract}
A positive muon spin rotation and relaxation ($\mu^+$SR) 
experiment on [Ca$_2$CoO$_3$]$_{0.62}$[CoO$_2$], ({\sl i.e.}, 
Ca$_3$Co$_4$O$_9$, a layered thermoelectric cobaltite) 
indicates the existence of two magnetic transitions at $\sim$ 100~K and 
400 - 600~K; the former is a transition from a paramagnetic state 
to an incommensurate ({\sf IC}) 
spin density wave ({\sf SDW}) state. 
The anisotropic behavior of 
zero-field $\mu^+$SR spectra at 5~K suggests that the {\sf IC-SDW} 
propagates in the $a$-$b$ plane, 
with oscillating moments directed along 
the $c$-axis; 
also the {\sf IC-SDW} is found to
exist not in the [Ca$_2$CoO$_3$] subsystem 
but in the [CoO$_2$] subsystem. 
In addition, it is found that 
the long-range {\sf IC-SDW} order completes below $\sim$ 30~K, 
whereas the short-range order appears below 100~K. 
The latter transition is interpreted as a gradual change in 
the spin state of Co ions 
above 400~K.  
These two magnetic transitions detected by $\mu^+$SR 
are found to correlate closely with 
the transport properties of 
[Ca$_2$CoO$_3$]$_{0.62}$[CoO$_2$].  
\end{abstract}


\maketitle

\section{\label{sec:Intro}Introduction}
A strong correlation between $3d$ electrons 
induces important physical properties in 3$d$ metal oxides; 
${\it e.g.}$ high temperature superconductivity in cuprates, 
colossal magnetoresistance in manganites and 
probably 'good' thermoelectric properties in layered cobaltites.  
Four cobaltites, 
[Ca$_2$CoO$_3$]$_{0.62}$[CoO$_2$],
\cite{CCO_1,CCO_2,CCO_3}
Na$_x$CoO$_2$ with $x \sim 0.6$,
\cite{NCO_1,NCO_2,NCO_3}
[Sr$_2$Bi$_{2-y}$Pb$_y$O$_4$]$_x$[CoO$_2$],
\cite{4LBiSrCO_1,4LBiSrCO_2,4LBiSrCO_3}
and
[Ca$_2$Co$_{4/3}$Cu$_{2/3}$O$_4$]$_{0.62}$[CoO$_2$],
\cite{4LCCCO_1}
are known to be good thermoelectrics because of their 
metallic conductivities and high thermoelectric powers, 
for reasons which are currently not fully understood. 
In order to find excellent thermoelectrics 
suitable for thermoelectric power generation 
for protecting the environment by saving energy resources 
and reducing the release of CO$_2$ into the atmosphere, 
it is crucial to understand the mechanism of 
the 'good' thermoelectric properties in 
these layered cobaltites.

The layered cobaltites 
share a common structural component: 
the CoO$_2$ planes, in which a two-dimensional-triangular lattice 
of Co ions is formed by a network of edge-sharing CoO$_6$ octahedra.  
Charge carrier transport in these materials is thought to be 
restricted mainly to these CoO$_2$ planes, 
as in the case of the CuO$_2$ planes for the high-$T_c$ cuprates.  
Since specific heat measurements on Na$_x$CoO$_2$ 
indicate a large thermal effective mass of carriers \cite{NCO_4}, 
all these cobaltites are believed to be 
strongly correlated electron systems.  

The crystal structure of 
[Ca$_2$CoO$_3$]$_{0.62}$[CoO$_2$] 
consists of alternating layers of 
the triple rocksalt-type [Ca$_2$CoO$_3$] subsystem 
and the single CdI$_2$-type [CoO$_2$] subsystem 
stacked along the $c$-axis \cite{CCO_2,CCO_3,CCO_4}.  
There is a misfit between these subsystems along the $b$-axis.  
Susceptibility ($\chi$) measurements \cite{CCO_2,jun_PRB1} 
indicate two magnetic transitions at 19~K and 380~K; 
the former is a ferrimagnetic transition ($T_{\sf FR}$) 
and the latter is probably a spin-state transition 
($T_{\sf SS}^{\chi})$.  
The temperature dependence of the resistivity $\rho$ 
exhibits a broad minimum around 80~K \cite{CCO_2,CCO_3,jun_PRB1} 
and a broad maximum between 400 and 600~K \cite{CCO_2}.  
Although $\rho$ appears to diverge with decreasing temperature 
below $T_{\sf FR}$, it is worth noting that $\chi(T)$ 
shows no clear anomalies near 80~K or 600~K.  

A recent positive muon spin rotation and relaxation 
($\mu^+$SR) experiment \cite{jun_PRB1,jun_PhysicaB1} 
indicated the existence of an incommensurate ({\sf IC}) 
spin density wave ({\sf SDW}) state below 100~K, 
which was not detected previously by other magnetic 
measurements \cite{CCO_2,CCO_3}. 
Thus, the broad minimum 
around 80~K in the $\rho(T)$ curve 
suggests that 
the behavior of conduction electrons is 
strongly affected by the {\sf IC-SDW} order
in [Ca$_2$CoO$_3$]$_{0.62}$[CoO$_2$].
Nevertheless, we need more information to confirm  
the correlation between the transport properties 
and the {\sf IC-SDW} in 
[Ca$_2$CoO$_3$]$_{0.62}$[CoO$_2$]; 
such as the structure of the {\sf IC-SDW} and 
the subsystem in which the {\sf IC-SDW} exists. 
Furthermore, $T_{\sf SS}^{\chi}(\sim 380$~K) is too low 
to explain the whole change in the $\rho(T)$ curve 
between 400 and 600~K,
while the $\mu^+$SR experiment showed 
a change in slope of the relaxation rate-{\sl vs.}-$T$ curve 
above 400~K.\cite{jun_PhysicaB1}
 
In order to further clarify the role of magnetism 
in thermoelectric layered cobaltites, 
we have measured both weak ($\sim$ 100~Oe) transverse-field 
positive muon spin rotation and relaxation (wTF-$\mu^+$SR) 
and zero field (ZF-) $\mu^+$SR time spectra 
in [Ca$_2$CoO$_3$]$_{0.62}$[CoO$_2$] at temperatures below 700~K.  
The former method is sensitive to local magnetic order 
{\it via\/} the shift of the $\mu^+$ spin precession frequency 
and the enhanced $\mu^+$ spin relaxation, 
while ZF-$\mu^+$SR is sensitive to weak local magnetic [dis]order 
in samples exhibiting quasi-static paramagnetic moments.  

\section{\label{sec:Expt}Experiment}
A randomly oriented polycrystalline disk 
($\sim 20$~mm diameter and $\sim 2$~mm thick) of 
[Ca$_2$CoO$_3$]$_{0.62}$[CoO$_2$] 
was synthesized by a conventional 
solid state reaction technique \cite{jun_PRB1}.  
$C$-axis aligned polycrystalline 
[Ca$_2$CoO$_3$]$_{0.62}$[CoO$_2$] and 
[Ca$_{1.8}M_{0.2}$CoO$_3$]$_{0.62}$[CoO$_2$]
($M$ = Sr, Y, Bi) plates 
($\sim 20 \times 20 \times 2$~mm$^3$) 
were synthesized by a reactive templated 
grain growth technique \cite{rtgg_1}.  
Single-crystal platelets of [Ca$_2$CoO$_3$]$_{0.62}$[CoO$_2$] 
($\sim 5 \times 5 \times 0.1$~mm$^3$) 
were prepared by a SrCl$_2$ flux method \cite{CCO_5}.  
Then, all the samples were annealed 
in an O$_2$ flow at 450~$^o$C for 12 hours.
The preparation and characterization of these samples 
were described in detail elsewhere \cite{rtgg_2,sc_1}.  
The $\mu$SR experiments were performed on the 
{\bf M20} and {\bf M15} surface muon beam line at TRIUMF.  
The experimental setup is described elsewhere \cite{ICSDW_1}.  

\section{\label{sec:Results}Results} 
\subsection{\label{ssec:ICSDW}{\sf IC-SDW} transition} 
\begin{figure}
\includegraphics[width=8cm]{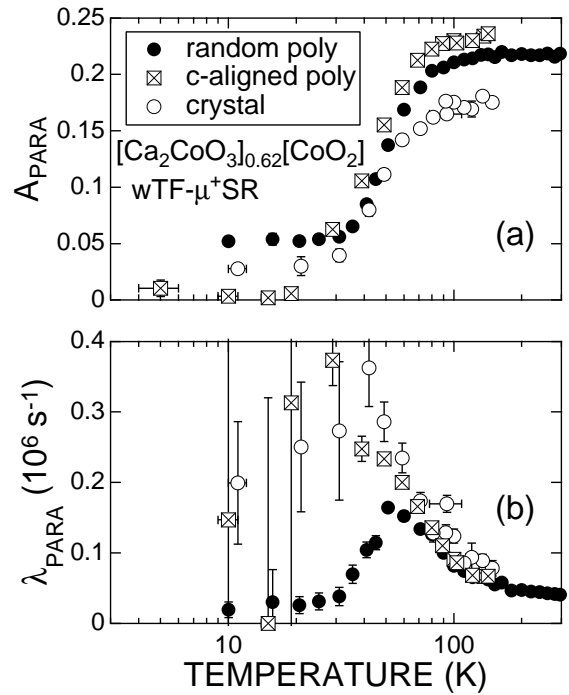} 
\caption{\label{fig:wTF-muSR} 
  (a) Paramagnetic $\mu^+$ spin precession asymmetry $A_{\sf PARA}$ and 
  (b) muon spin relaxation rate $\lambda_{\sf PARA}$ 
  as a function of temperature for the three 
  [Ca$_2$CoO$_3$]$_{0.62}$[CoO$_2$] samples: 
  a randomly oriented polycrystalline disk (solid circles) \cite{jun_PRB1}, 
  a $c$-axis aligned polycrystalline plate (squares) 
  and single crystal ({\sf sc}) platelets (open circles).  
  For the {\sf sc} platelets, both the value of $A_{\sf PARA}$ above 100~K
  and the change in $A_{\sf PARA}$ below 100~K
  are smaller than those in the polycrystalline samples.
  This is because the muon momentum was decreased from 28 to 25~MeV/c
  for the {\sf sc} measurements to stop muons in the thin platelets
  ($\sim 100$~$\mu$m thickness), causing a small background signal
  from muons stopping elsewhere.
}
\end{figure}

In all the [Ca$_2$CoO$_3$]$_{0.62}$[CoO$_2$] samples, 
the wTF-$\mu^+$SR spectra in a magnetic field of $H \sim100$~Oe 
exhibit a clear reduction of the $\mu^+$ precession amplitude below 100~K.  
The data were obtained by fitting the wTF-$\mu^+$SR spectrum
in the time domain with a combination of
a slowly relaxing precessing signal and
two non-oscillatory signals, one fast and the other slow relaxing:
\begin{eqnarray}
A_0 \, P(t) &=& A_{\sf PARA} \, \exp(- \lambda_{\sf PARA} t) \, \cos (\omega_\mu t + \phi)
\cr
&+& A_{\sf fast} \, \exp(-\lambda_{\sf fast} t) 
\cr
&+& A_{\sf slow} \, \exp(-\lambda_{\sf slow} t), 
\label{eq:TFfit}
\end{eqnarray}
where $A_0$ is the initial asymmetry, 
$P(t)$ is the muon spin polarization function, 
$\omega_\mu$ is the muon Larmor frequency, 
$\phi$ is the initial phase of the precession and 
$A_n$ and $\lambda_n$ ($n$ = {\sf PARA}, {\sf fast} and {\sf slow}) 
are the asymmetries and exponential relaxation rates of the three signals.  
The latter two signals ($n$ = {\sf fast} and {\sf slow})
have finite amplitudes
below $T_{\sf SDW}^{\rm on} \approx 100$~K
and probably suggest the existence of multiple muon sites
in [Ca$_2$CoO$_3$]$_{0.62}$[CoO$_2$].

Figures \ref{fig:wTF-muSR}(a) and \ref{fig:wTF-muSR}(b) 
show the temperature dependences of 
the paramagnetic asymmetry $A_{\sf PARA}$ 
(which is proportional to the volume fraction of 
a paramagnetic phase in the sample) 
and the corresponding relaxation rate $\lambda_{\sf PARA}$ 
in three [Ca$_2$CoO$_3$]$_{0.62}$[CoO$_2$] samples:
a randomly oriented polycrystalline sample \cite{jun_PRB1},
a $c$-aligned polycrystalline sample,
and single crystal platelets.
The large decrease in $A_{\sf PARA}$ below 100~K 
(and the accompanying increase in $\lambda_{\sf PARA}$) 
indicate the existence of a magnetic transition 
with an onset temperature $T_c^{\rm on} \approx 100$~K 
and a transition width $\Delta T \approx 70$~K.  
The single crystal data suggest that the large $\Delta T$ 
is not caused by inhomogeneity of the sample 
but is an intrinsic property of this compound.  

\begin{figure}
\includegraphics[width=8cm]{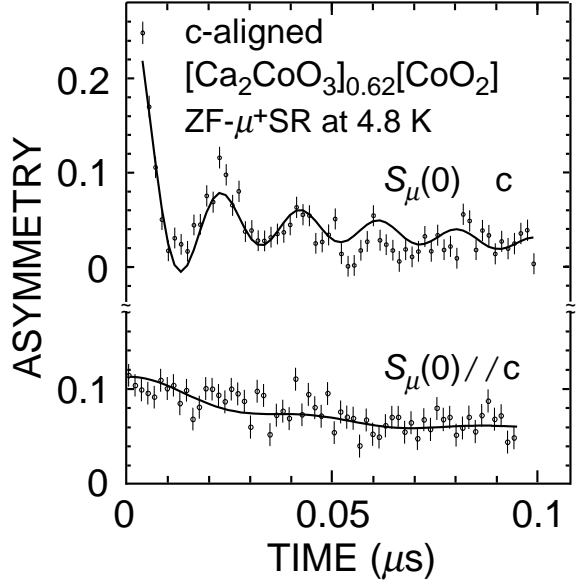} 
\caption{\label{fig:ZF-muSR} 
  ZF-$\mu^+$SR time spectra of 
  the $c$-aligned [Ca$_2$CoO$_3$]$_{0.62}$[CoO$_2$] plate at 4.8~K. 
  The configurations of the sample and the initial muon spin direction 
  $\vec{\bm S}_\mu(0)$ are 
  (top) $\vec{\bm S}_\mu(0) \perp \hat{\bm c}$ and 
  (bottom) $\vec{\bm S}_\mu(0) \parallel \hat{\bm c}$.  
  }
\end{figure}

Figure \ref{fig:ZF-muSR} shows ZF-$\mu^+$SR time spectra at 4.8~K 
in the $c$-aligned sample; the top spectrum was obtained with 
the initial $\mu^+$ spin direction $\vec{\bm S}_\mu(0)$ 
perpendicular to the $c$-axis and the bottom one with 
$\vec{\bm S}_\mu(0) \parallel \hat{\bm c}$.  
A clear oscillation due to quasi-static internal fields 
is observed only when $\vec{\bm S}_\mu(0) \perp \hat{\bm c}$. 
The time interval from $t=0$ to the first zero crossing 
of that oscillation is roughly the same ($1:1.2954$) 
as the interval between the first and second zero crossings; 
this is a characteristic of a zeroth-order Bessel function of the first kind $J_0(\omega_\mu t)$ 
that describes the muon polarization evolution 
in an incommensurate spin density wave {\sf IC-SDW} field distribution 
\cite{ICSDW_1,ICSDW_2,ICSDW_3}.

Actually, the top oscillating spectrum was fitted
using a combination of three signals: 
\begin{eqnarray}
 A_0 \, P(t) &=& 
   A_{\sf SDW} \, J_0(\omega_{\mu} t) \, \exp(-\lambda_{\sf SDW} t)
\cr
 &+& A_{\rm KT} \, G_{zz}^{\rm KT}(t,\Delta) 
\cr
 &+& A_{\rm tail} \, \exp(-\lambda_{\rm tail} t),
\label{eq:ZFfit}
\end{eqnarray}
\begin{equation}
 \omega_\mu \equiv  2 \pi \nu_\mu = \gamma_{\mu} \; H_{\sf int},
\label{eq:omg}
\end{equation}
\begin{eqnarray}
 &~& G_{zz}^{\rm KT}(t,\Delta) \; = \; {1\over3} 
\cr
 &+& {2\over3} \, \left( 1 - \Delta^2 t^2 \right) 
 \exp(- \Delta^2 t^2 / 2),
\label{eq:GKT}
\end{eqnarray}
where $A_0$ is the empirical maximum muon decay asymmetry, 
$A_{\sf SDW}$, $A_{\rm KT}$ and $A_{\rm tail}$ 
are the asymmetries associated with the three signals, 
$G_{zz}^{\rm KT}(t,\Delta)$ is the static Gaussian Kubo-Toyabe function, 
$\Delta$ is the static width of the distribution of local frequencies 
at the disordered sites and 
$\lambda_{\rm tail}$ is the slow relaxation rate of the 'tail' 
(not shown in this Figure),
and the fit using an exponential relaxed cosine oscillation,
$\exp(-\lambda t) \cos(\omega_{\mu} t + \phi)$,
provides a phase angle $\phi \sim 90^{\rm o}$, 
which is physically meaningless.\cite{ICSDW_4}

We therefore conclude that [Ca$_2$CoO$_3$]$_{0.62}$[CoO$_2$] undergoes 
a magnetic transition from a paramagnetic state to an {\sf IC-SDW} state 
({\it i.e.} $T_c^{\rm on} = T_{\sf SDW}^{\rm on}$).  
The absence of a clear oscillation in the bottom spectrum 
of Fig.~\ref{fig:ZF-muSR} indicates that 
the internal magnetic field $\vec{\bm H}_{\rm int}$ 
is roughly parallel to the $c$-axis, since the muon spins 
do not precess in a parallel magnetic field.  
The {\sf IC-SDW} is unlikely to propagate along the $c$-axis 
due both to the two-dimensionality 
and to the misfit between the two subsystems.  
The {\sf IC-SDW} is therefore considered to propagate in the $a$-$b$ plane, 
with oscillating moments directed along the $c$-axis.  
This suggests that the ferrimagnetic interaction 
is also parallel to the $c$-axis,
and is consistent with the results of 
our $\chi$ measurement on single crystals.\cite{jun_PRB2}  

\begin{figure}
\includegraphics[width=8cm]{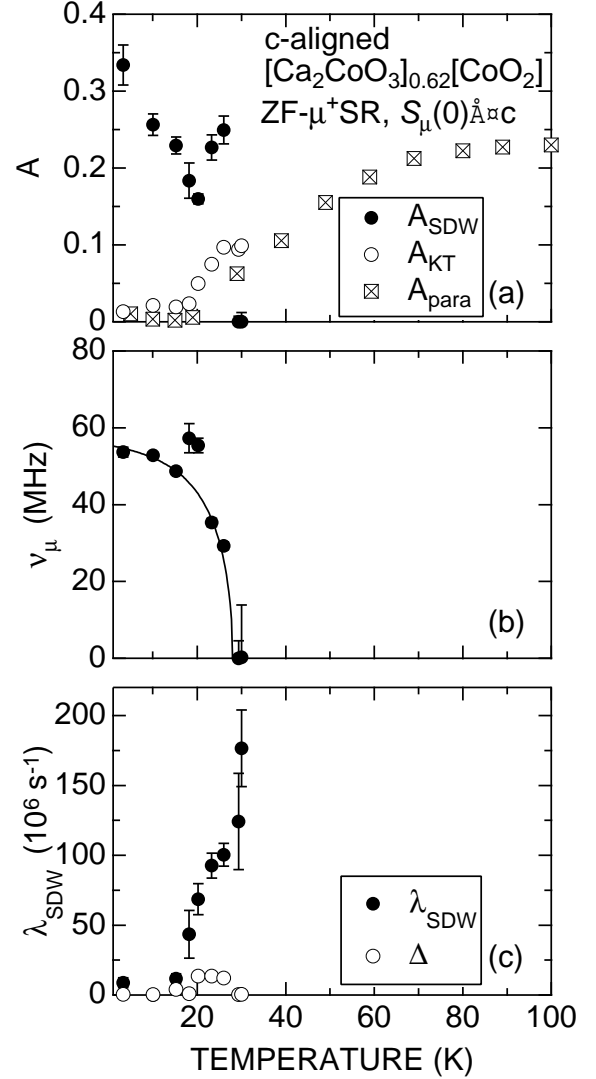}
\caption{\label{fig:SDW_3L} 
 Temperature dependences of 
 (a) $A_{\sf SDW}$ and $A_{\sf PARA}$ 
 (estimated by the wTF-$\mu^+$SR experiment), 
 (b) $\nu_{\mu}$ and 
 (c) $\lambda_{\sf SDW}$ 
 fot the $c$-aligned 
 [Ca$_2$CoO$_3$]$_{0.62}^{\rm RS}$[CoO$_2$]. 
 The solid line in Fig.~\ref{fig:SDW_3L}(b) represents 
 the temperature dependence of the {\sf BCS} gap energy. 
 The deviation of the experimental data  
 from the theory around 20~K is probably due to the effect of 
 the ferrimagnetic transition at 19~K.
 }
\end{figure}
Figures~\ref{fig:SDW_3L}(a)-\ref{fig:SDW_3L}(c) 
show the temperature dependences of 
$A_{\sf SDW}$, $A_{\sf KT}$ and $A_{\sf PARA}$ 
(same in Fig.~\ref{fig:wTF-muSR}),
$\nu _{\mu}$ and $\lambda _{\sf SDW}$ and $\Delta$ 
for the $c$-aligned 
[Ca$_2$CoO$_3$]$_{0.62}^{\rm RS}$[CoO$_2$].
$A_{\sf SDW}$ increases 
with decreasing $T$ below 30~K, 
although $A_{\sf PARA}$ obtained by the wTF-$\mu^+$SR measurement 
exhibits a rapid decrease below 100~K and levels off to almost 0 below 30~K 
(see Fig.~\ref{fig:SDW_3L}(a)). 
According to the recent $\chi$ measurements 
using single crystal platelets,\cite{jun_PRB2} 
a small shoulder in the $\chi (T)$ curve was observed at 27~K 
only for $H \bot ab$. 
This temperature (27~K) corresponds to the highest temperature 
that a clear $\mu^+$SR signal 
due to the {\sf IC-SDW} was observed. 
Thus, it is considered that a short-range order {\sf IC-SDW} state 
appears below 100~K = $T_{\rm SDW}^{\rm on}$, 
while the long-range order is completed below 27~K; {\sl i.e.}, 
$T_{\rm SDW}$ = $T_{\rm SDW}^{\rm end}$. 
Since both $\rho (T)$ and $S(T)$ are metallic above 80~K 
and semiconducting below 80~K,\cite{CCO_2,CCO_3}
charge carrier transport is strongly affected by 
a formation of the short-range {\sf IC-SDW} order.

Although the $\nu _{\mu}(T)$ curve 
is well explained by the {\sf BCS} weak coupling theory 
as expected for the {\sf IC-SDW} state,\cite{ICSDW_5}  
there is a deviation  
from the theory around 20~K
(see Fig.~\ref{fig:SDW_3L}(b)).
This deviation (and the accompanying increase in $A_{\sf SDW}$)
is probably due to the effect of 
the ferrimagnetic transition at 19~K (= $T_{\rm FR}$).
Here, the ferrimagnetism is considered to be caused 
by an interlayer coupling 
between Co spins in the [Ca$_2$CoO$_3$] and [CoO$_2$]
subsystems,\cite{jun_PRB2} 
while the {\sf IC-SDW} order completes below 27~K.
This means that the {\sf IC-SDW} is affected
by the ferrimagnetic coupling
{\sl via.} the Co spins in the [Ca$_2$CoO$_3$] subsystem.
Therefore, the enhancement of the internal magnetic field 
at $T_{\rm FR}$
is likely to be caused by a critical phenomenon
around the ferrimagnetic transition.
In addition, 
the magnitude of $\lambda_{\sf SDW}$ 
decreases rapidly with decreasing $T$ and levels off to 
a constant value below 20~K. 
This suggests that 
the broadening of the {\sf IC-SDW} field distribution 
at the $\mu^+$ sites
mainly occurs in the temperature range 
between $T_{\rm SDW}$ and $T_{\rm FR}$.

\begin{figure}
\includegraphics[width=8cm]{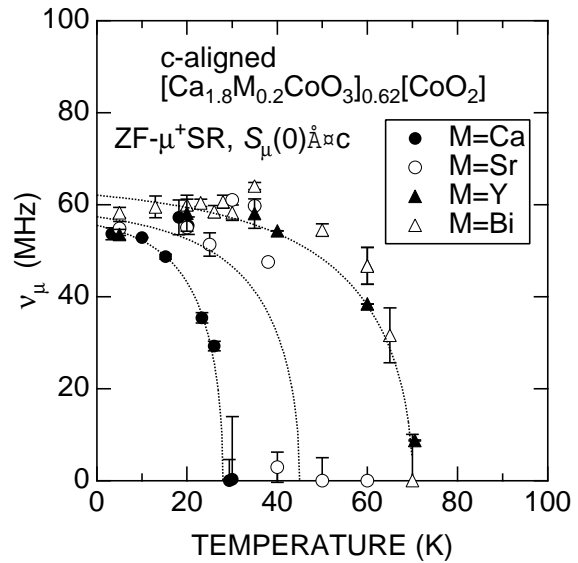}
\caption{\label{fig:SDW_others} 
 Temperature dependences of $\nu _{\mu}$ for 
 the $c$-aligned pure and doped  
 [Ca$_2$CoO$_3$]$_{0.62}^{\rm RS}$[CoO$_2$].
 The dotted lines represent 
 the temperature dependence of the {\sf BCS} gap energy. 
  }
\end{figure}

In order to determine the subsystem in which the {\sf IC-SDW} exists, 
ZF-$\mu^+$SR spectra were measured in doped samples: 
$c$-aligned polycrystalline 
[Ca$_{1.8}M_{0.2}$CoO$_3$]$_{0.62}$[CoO$_2$]
($M$ = Sr, Y and Bi).
A clear precession was observed 
in the ZF-$\mu^+$SR spectrum 
with $\vec{\bm S}_\mu(0) \perp \hat{\bm c}$
in every sample,
although $T_{\rm SDW}$ depended on dopant.
Figure~\ref{fig:SDW_others} shows the temperature dependences of
$\nu _{\mu}$ for the $c$-aligned pure and doped  
[Ca$_2$CoO$_3$]$_{0.62}^{\rm RS}$[CoO$_2$] samples.
Doping with Y and Bi increase $T_{\sf SDW}$ by $\sim$ 40~K 
and Sr-doping by $\sim$ 20~K, 
although Sr-doping did not affect $T_{\sf SDW}^{\rm on}$ 
by the previous wTF-$\mu^+$SR experiment.\cite{jun_PRB1}   

It should be noted that
all the samples show approximately the same 
precession frequency ($\sim 60$~MHz) at zero temperature.  
This suggests that the local magnetic field $H_{\rm int}$(0~K) 
is independent of dopant.  
Since $H_{\rm int}$ in the doped [Ca$_2$CoO$_3$] subsystem 
should be strongly affected by the dopant, it is concluded that 
the {\sf IC-SDW} exists not in the [Ca$_2$CoO$_3$] subsystem 
but in the [CoO$_2$] subsystem. 
Also, the latest $\mu^+$SR experiment on 
[Ca$_2$Co$_{4/3}$Cu$_{2/3}$O$_4$]$_{0.62}$[CoO$_2$],
\cite{jun_PRB4} 
which consists of the quadruple rocksalt-type subsystem 
and the single [CoO$_2$] subsystem, 
also indicates the existence of an {\sf IC-SDW} state 
below $\sim 200$~K. 
The precession frequency due to an internal {\sf IC-SDW} field 
is estimated as $\sim 60$~MHz at zero temperature. 
This strongly suggests that the {\sf IC-SDW} exists 
in the [CoO$_2$] subsystem, 
because one third of the Co ions in the rocksalt-type subsystem 
are replaced by Cu ions. 
Therefore, the {\sf IC-SDW} is found to be caused 
by the spin-order of the conduction electrons 
in the [CoO$_2$] subsystem. 

\subsection{\label{ssec:SST} Spin State Transition} 
\begin{figure}
\includegraphics[width=7.8cm]{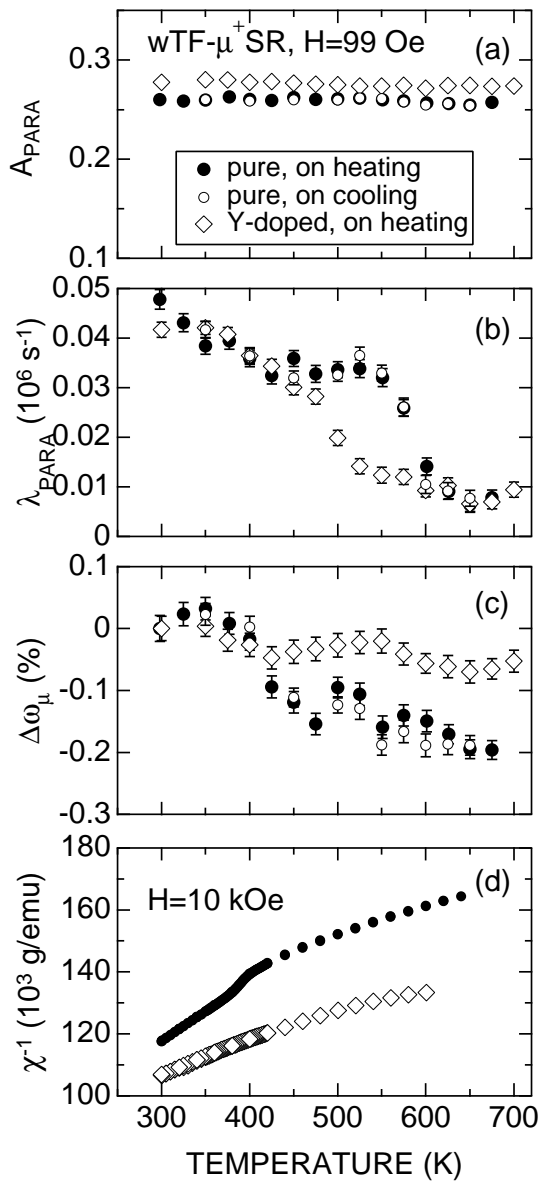} 
\caption{\label{fig:highT} 
  Temperature dependences of 
  (a) the asymmetry $A_{\sf PARA}$
  (b) the muon spin relaxation rate $\lambda_{\sf PARA}$  
  (c) the shift of the muon precession frequency $\Delta\omega_{\mu}$ and
  (d) the inverse susceptibility $\chi^{-1}$ 
  in a $c$-aligned polycrystalline 
  [Ca$_2$CoO$_3$]$_{0.62}$[CoO$_2$] sample (circles) 
  and a polycrystalline 
  [Ca$_{1.8}$Y$_{0.2}$CoO$_3$]$_{0.62}$[CoO$_2$]  
  sample (diamonds); 
  $A_{\sf PARA}$ and $\lambda_{\sf PARA}$ was obtained 
  by fitting the wTF-$\mu^+$SR spectrum in the time domain 
  using a simple exponential relaxation function, 
  $A_{\sf PARA}\, \exp(-\lambda_{\sf PARA} t)\, \cos(\omega_{\mu} t + \phi)$.  
  }
\end{figure}
The high-temperature wTF-$\mu^+$SR spectra were 
measured in an air flow 
to avoid the formation of oxygen deficiency
in the sample,
whereas the previous experiment in vacuum.
\cite{jun_PhysicaB1}
The spectra in the $c$-aligned 
[Ca$_2$CoO$_3$]$_{0.62}$[CoO$_2$] 
sample were well fitted using an exponential relaxed cosine oscillation,
$A_{\sf PARA} \exp(-\lambda_{\sf PARA} t) \cos(\omega_{\mu} t + \phi)$.
Figures~\ref{fig:highT}(a) - \ref{fig:highT}(d) show
the temperature dependences of $A_{\sf PARA}$, $\lambda_{\sf PARA}$, 
the shift of $\omega_{\mu}$ ($\Delta\omega_{\mu}$) and
the inverse susceptibility $\chi^{-1}$ 
in the $c$-aligned polycrystalline 
[Ca$_2$CoO$_3$]$_{0.62}$[CoO$_2$] sample  
and a polycrystalline 
[Ca$_{1.8}$Y$_{0.2}$CoO$_3$]$_{0.62}$[CoO$_2$]  
sample.  
Here, $\Delta\omega_{\mu}$ is defined as
($\omega_{\mu}(T)$-$\omega_{\mu}$(300~K))/$\omega_{\mu}$(300~K);
since the oscillation of a reference was not measured, 
$\Delta\omega_{\mu}$ is inequivalent to the muonic Knight shift.

A broad shoulder is clearly seen  
in the $\lambda_{\sf PARA}(T)$ curve 
of the pure sample at 400 - 600~K, 
although such a shoulder seems to be ambiguous 
in the Y-doped sample [Fig.~\ref{fig:highT}(b)].  
Moreover, as $T$ increases, 
the $\Delta\omega_{\mu}(T)$ curve exhibits a sudden decrease 
at $\sim$ 400~K, 
while the $\Delta\omega_{\mu}(T)$ curve in the Y-doped sample
is roughly independent of $T$.
It should be noted that, as seen in Figs.~\ref{fig:wTF-muSR}(a) 
and \ref{fig:highT}(a), above 150~K 
$A_{\sf PARA}$ levels off to its maximum value ($\sim 0.26$) 
--- {\it i.e.} the sample volume is almost 100\% paramagnetic.  
In addition, there is no thermal hysteresis in the data 
for the $c$-aligned  
[Ca$_2$CoO$_3$]$_{0.62}$[CoO$_2$] sample 
obtained on heating and on cooling. 
This suggests that the changes in the $\lambda_{\sf PARA}$
and the $\Delta\omega_{\mu}$ are not caused by the formation
of oxygen deficiency but by a magnetic transition,
as discussed later.

These behaviors are in good agreement with 
the results of $\chi (T)$ measurements.  
That is, the $\chi^{-1}(T)$ curve of the pure sample 
exhibits an obvious change in slope 
at $T_{\sf SS}^{\chi} = 380$~K, 
while that of the Y-doped sample does not 
[Fig.~\ref{fig:highT}(d)].  
The change in the $\chi^{-1}(T)$ curve is 
considered to be 
attributed to the spin state transition 
of the Co$^{3+}$ and Co$^{4+}$ ions 
from the low temperature $LS$ or $LS+IS$ 
to the high-temperature $LS+IS$, $IS$, $IS+HS$ or $HS$,
\cite{CCO_2,jun_PRB2} 
as in the case of LaCoO$_3$.\cite{LCO_1,LCO_2} 
Here $LS$, $IS$ and $HS$ are the low-spin 
($t_{2g}^6$; $S$=0 and $t_{2g}^5$; $S$=1/2), intermediate-spin 
($t_{2g}^5e_g^1$; $S$=1 and $t_{2g}^4e_g^1$; $S$=3/2) and high-spin 
($t_{2g}^4e_g^2$; $S$=2 and $t_{2g}^3e_g^2$; $S$=5/2) states, respectively.  

At these temperatures muons are diffusing rapidly, 
so that the relaxation rate usually decreases monotonically 
with increasing temperature.  
Hence we can conclude that 
both the shoulder in the $\lambda_{\sf PARA}(T)$ curve 
and the sudden decrease in the $\Delta\omega_{\mu}(T)$ curve
are induced by the spin state transition, 
because there is no indications for the appearance 
of a magnetically ordered state 
(see Fig.~\ref{fig:highT}(a)). 
Therefore, the spin state transition 
from the low-temperature $LS$ 
to the high-temperature $IS+HS$ or $HS$ 
is most reasonable 
to explain the change in $H_{\rm int}$ 
(suggested by the changes in 
$\lambda_{\sf PARA}(T)$ and $\Delta\omega_{\mu}(T)$)
without the magnetic order, {\sl i.e.},
the temperature independent $A_{\sf PARA}(T)$.
On the other hand, both the rapid muon diffusion 
and the fast exchange rate of electrons 
between Co$^{3+}$ and Co$^{4+}$ ions 
decrease $\lambda_{\sf PARA}$ with increasing $T$.  
The competition between these three factors 
is likely responsible for the broad shoulder 
in $\lambda_{\sf PARA}(T)$ around 400 - 600~K.  

\begin{figure}
\includegraphics[width=8cm]{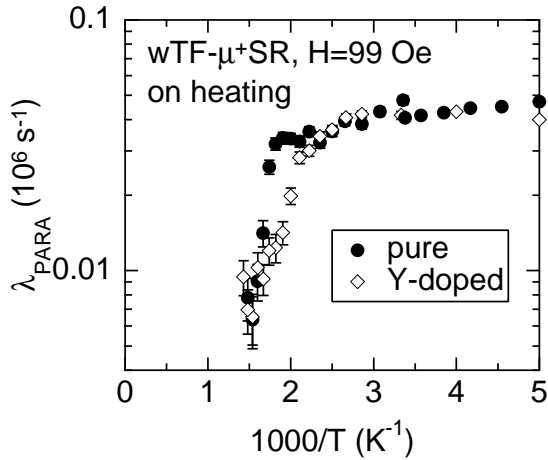} 
\caption{\label{fig:highTI} 
  Muon spin relaxation rate $\lambda_{\sf PARA}$  
  as a function of $T^{-1}$ in a $c$-aligned polycrystalline 
  [Ca$_2$CoO$_3$]$_{0.62}$[CoO$_2$] sample (circles) 
  and a polycrystalline 
  [Ca$_{1.8}$Y$_{0.2}$CoO$_3$]$_{0.62}$[CoO$_2$]  
  sample (diamonds).  
  A discontinuous in data at 300~K was caused by the change 
  in the experimental setup from a cryostat to an oven.
  }
\end{figure}
In order to know the contribution from the latter two factors, 
Fig.~\ref{fig:highTI} shows the relationship 
between $\lambda_{\sf PARA}$ and $T^{-1}$ 
of the pure and Y-doped samples, 
because the latter two factors are expected to depend on $\exp(T^{-1})$.
Nevertheless, the linear relationship is not observed 
even in the Y-doped sample; 
thus, it is difficult to separate the contribution from each factor 
at present, although the difference between both samples 
are clearly seen in Fig.~\ref{fig:highTI}. 
Indeed, the $\lambda_{\sf PARA}(T)$ curves 
of the pure and Y-doped samples 
seem to level off to a constant value 
($\sim 0.01 \times$ 10$^6$ sec$^{-1}$) above 650~K
due to a rapid muon diffusion,
as in the case of YBa$_2$Cu$_3$O$_{6\pm\delta}$.
\cite{YBCO_1,YBCO_2}
Therefore, we can not determine the onset temperature 
($T_{\sf SS}^{\rm on}$)
of the broad shoulder in the $\lambda_{\sf PARA}(T)$ curve , 
based only on the present $\mu^+$SR result,
although $T_{\sf SS}^{\rm on} \geq 600$~K.
The broad shoulder also suggests the possibility 
that the spin state changes gradually 
above 400~K.  
In other words, $T_{\sf SS}^{\rm on} \geq 600$~K 
and the endpoint $T_{\sf SS}^{\rm end} = T_{\sf SS}^{\chi} = 380$~K. 
And at temperatures between 
$T_{\sf SS}^{\rm on}$ and $T_{\sf SS}^{\rm end}$,
the populations of the $IS$ and $HS$ states 
are likely to vary as a function of temperature,
as in the case of LaCoO$_3$.\cite{LCO_1,LCO_2}  
The relationship between the spin state transition and 
the transport properties is discussed later. 

\section{\label{sec:Discussion}Discussion} 
\subsection{\label{ssec:ICSDW1} The nature of {\sf IC-SDW}}
%
%
%
%

There are two Co sites in the 
[Ca$_2$CoO$_3$]$_{0.62}^{\rm RS}$[CoO$_2$] 
lattice; 
thus, it is difficult to determine the Co valence in the [CoO$_2$] 
plane by a $\chi$ measurement or a chemical titration technique, 
although both Co$^{3+}$ and Co$^{4+}$ ions 
are mainly in the $LS$ state
below $T_{\sf SS}^{\rm end}$,
according to the photo-emission and x-ray absorption studies on 
the related cobaltiets, 
[Sr$_2$Bi$_{2-y}$Pb$_y$O$_4$]$_x$[CoO$_2$]
\cite{4LBiSrCO_4}
and $\chi$ measurements on several cobaltites.
\cite{NCO_8,CCO_2,CCO_3,jun_PRB1,4LBiSrCO_1} 
If we assume that only Co$^{3+}$ and Co$^{4+}$ ions exist in 
[Ca$_2$CoO$_3$]$_{0.62}^{\rm RS}$[CoO$_2$] 
the average valence of the Co ions 
in the [CoO$_2$] plane is calculated as +3.38. 
This value is similar to the nominal valence of Co ions 
in Na$_{0.6}$CoO$_2$.
Hence, the Co spins with $S$ = 1/2 are considered to occupy 
40\% corners in the two dimensional triangular lattice 
to minimize an electric repulsion and a geometrical frustration
in the {\sf IC-SDW} state. 

It is worth noting that the $\mu^+$ sites are bound to the O ions
in the [CoO$_2$] plane. 
This means that the $\mu^+$ mainly feel the magnetic field 
in the [CoO$_2$] plane. 
Thus, the {\sf IC-SDW} is most unlikely to be caused 
by the misfit between the two subsystems, 
but to be an intrinsic behavior of the [CoO$_2$] plane. 
Nevertheless, the structure of the {\sf IC-SDW} order is 
still unknown at present, 
because the current $\mu^+$SR experiments provide 
no information 
on the incommensurate wave vector.
In order to obtain such information, 
it is necessary to carry out 
both $^{59}$Co-NMR and neutron scattering experiments
on these cobaltites.

\subsection{\label{ssec:ICSDW2} {Calculation and other experiments on 
\sf IC-SDW}}

The {\sf IC-SDW} order in 
[Ca$_2$CoO$_3$]$_{0.62}^{\rm RS}$[CoO$_2$]
is assigned to be a spin ($S$=1/2) order 
on the two-dimensional triangular lattice at non-half filling.
Such a problem was investigated by several workers 
using the Hubbard model within a mean field approximation;
\cite{MHonTL_1,MHonTL_2,MHonTL_3,MHonTL_4}
\begin{eqnarray}
 H&=&-t\sum_{<ij>\sigma}c_{i\sigma}^{\dagger}c_{j\sigma} + 
 U\sum_i n_{i\uparrow}n_{i\downarrow} ,
\label{eq:Hubbard}
\end{eqnarray}
where $c_{i\sigma}^{\dagger}(c_{j\sigma})$ creates (destroys) 
an electron with spin $\sigma$ on site $i$, 
$n_{i\sigma}=c_{i\sigma}^{\dagger}c_{i\sigma}$ is the number operator, 
$t$ is the nearest-neighbor hopping amplitude and 
$U$ is the Hubbard on-site repulsion.
The electron filling $n$ is defined as $n$ = (1/2$N$)$\sum_i^N n_i$,
where $N$ is the total number of sites.  
At $T$=0 and $n$=0.5 ({\sl i.e.}, 
the average valence of the Co ions in the [CoO$_2$] plane is +4),
as $U$ increased from 0, the system is 
paramagnetic metal up to $U/t \sim 3.97$, 
and changes into a metal 
with a spiral {\sf IC-SDW}, 
and then at $U/t \sim 5.27$, 
a first-order metal-insulator transition occurs.\cite{MHonTL_1}
On the other hand, at $n$=0.81 ({\sl i.e.}, 
the average valence of the Co ions in the [CoO$_2$] plane is +3.38),
a spiral {\sf SDW} state is stable below $U/t \sim 3$.\cite{MHonTL_2} 
These calculations suggest that the {\sf IC-SDW} state is stable 
for a weak-to-moderate coupling ($U/t \leq 5$).
Also, the {\sf IC-SDW} phase boundary was reported to depend on $n$;
that is, in the $n$ range between 0.75 and 1, 
the largest $U/t$ for the {\sf IC-SDW} phase increased 
monotonically with increasing $n$.\cite{MHonTL_2}
Since larger $U/t$ stabilize the energy gap 
at higher temperature,\cite{MHonTL_4} 
the calculations are consistent with the present $\mu^+$SR results,
{\sl i.e.}, the large increase in $T_{\sf SDW}$ due to the Y- or Bi-doping into 
[Ca$_2$CoO$_3$]$_{0.62}^{\rm RS}$[CoO$_2$].

On the other hand, 
our preliminary heat capacity measurements using 
single crystal and $c$-aligned polycrystal 
[Ca$_2$CoO$_3$]$_{0.62}^{\rm RS}$[CoO$_2$] 
samples indicate that 
the electronic specific heat parameter $\gamma$ 
ranges from 60 to 90~mJK$^{-2}$ per mol [CoO$_2$],
if we ignore the magnetic contribution.
This value is higher than that for Na$_x$CoO$_2$ with $x \sim 0.5$
($\gamma \sim$ 24~mJK$^{-2}$ per mol Co).\cite{NCO_4} 
Thus, [Ca$_2$CoO$_3$]$_{0.62}^{\rm RS}$[CoO$_2$] 
seems to be a strongly correlated electron system,
as well as Na$_x$CoO$_2$. 
As a result, it is expected that $U \gg t$,
although the above calculations suggest $U/t \leq 5$. 
In order to solve this puzzle, 
it is necessary to determine the {\sf IC-SDW} structure
at first.
  
According to the recent photoelectron spectroscopic study 
on [Ca$_2$CoO$_3$]$_{0.62}^{\rm RS}$[CoO$_2$]
below ambient temperature, 
the density of states {\sf DOS} at the Fermi level $E_{\rm F}$ 
decreased with decreasing $T$ and 
disappeared at 10~K.\cite{CCO_UPS}
Thus, the energy gap was clearly observed at 10~K. 
This also supports our conclusion; that is, 
both the broad minimum at $\sim$ 80~K in the $\rho(T)$ curve 
and the rapid increase in $\rho$ below 80~K with decreasing $T$ 
are caused by the {\sf IC-SDW} order 
between the spins of the conduction electrons 
and not by a magnetic scattering,
such as, the Kondo effect.

\subsection{\label{ssec:SST} Effect of Oxygen Deficiency on Spin State Transition}
\begin{figure}
\includegraphics[width=8cm]{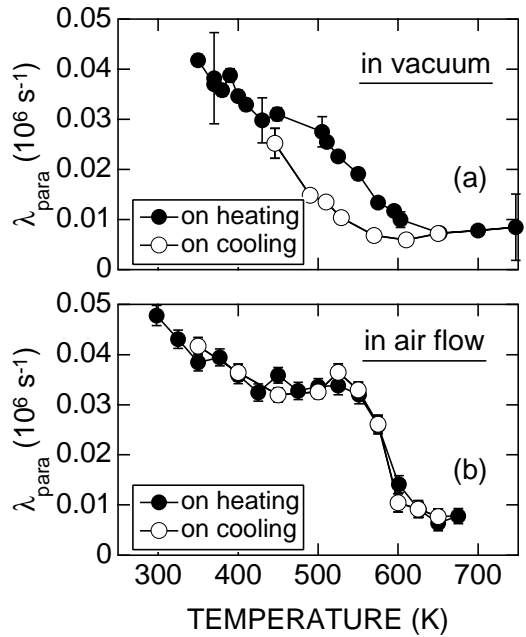} 
\caption{\label{fig:highT2} 
  Temperature dependences of 
  the muon spin relaxation rate $\lambda_{\sf PARA}$  
  for the [Ca$_2$CoO$_3$]$_{0.62}$[CoO$_2$] sample 
  measured 
  (a) in vacuum\cite{jun_PhysicaB1} and
  (b) in an air flow.  
  }
\end{figure}
In order to know the effect of atmosphere during the measurement,
Fig.~\ref{fig:highT2}
shows the comparison of the previous\cite{jun_PhysicaB1} 
and the present $\lambda_{\sf PARA}(T)$ curve; 
the former was measured in vacuum and the latter in an air flow.
There is a clear thermal hysteresis 
in the $\lambda_{\sf PARA}(T)$ curve
obtained in vacuum. 
In addition, the broad maximum between 400 and 600~K 
looks very ambiguous in the data obtained in vacuum 
compared with that in an air flow.
Recently, it was reported that oxygens are removed from 
[Ca$_2$CoO$_3$]$_{0.62}$[CoO$_2$] 
above 723~K (= 450~$^o$C) even in an oxygen flow.\cite{CCO_7}
Therefore, both the clear thermal hysteresis and 
the suppression of the broad maximum 
in the $\lambda_{\sf PARA}(T)$ curve 
obtained in vacuum 
are induced by the formation of oxygen deficiency.
As well as the oxygen deficiency, 
the substitution of Ca by Y decreases 
the average Co valence;\cite{jun_PRB1} 
as a result, the $\lambda_{\sf PARA}(T)$ curve
of the Y-doped sample looks quite similar to 
that of [Ca$_2$CoO$_3$]$_{0.62}$[CoO$_2$] 
obtained in vacuum on cooling
(see Fig.~\ref{fig:highT}(b) and Fig.~\ref{fig:highT2}(a)).

\subsection{\label{ssec:SST} Relationship between Spin State Transition 
and Transport Properties}
The broad shoulder in the $\lambda_{\sf PARA}(T)$ curve at 400 - 600~K
is in good agreement with the behavior of the $\rho (T)$ curve, 
because the $\rho (T)$ curve shows a broad maximum 
between 400 and 600~K, and above 600~K 
$\rho$ decreases monotonically with increasing $T$ 
up to 1000~K.\cite{CCO_2,CCO_6} 
On the other hand, 
the $S(T)$ curve exhibits a small change 
in slope around $T_{\sf SS}^{\chi}$;
that is, as $T$ increases from 0~K, 
$S$ increases monotonically up to $\sim$ 100~K and 
seems to level off a constant value 
($\sim 130~\mu$VK$^{-1}$) 
up to $\sim$ 400~K,
then $S$ again increases linearly 
(d$S$/d$T \sim$ 80~nVK$^{-2}$) up to 1000~K.\cite{CCO_2,CCO_6}
Therefore, above $T_{\sf SS}^{\rm end}$, 
the $\rho(T)$ curve exhibits a semiconducting behavior, 
whereas the $S(T)$ curve metallic.

The two Co sites in the 
[Ca$_2$CoO$_3$]$_{0.62}^{\rm RS}$[CoO$_2$] 
lattice leads to a question which is responsible 
for the spin state transition, 
as in the case of {\sf IC-SDW} order.
Both the change in slope of $S(T)$ at $\sim$ 400~K and 
the broad maximum of $\rho (T)$ at 400 - 600~K
suggest that the Co ions in the [CoO$_2$] plane 
play a significant role on the spin state transition.
Indeed, the related cobaltites, Na$_{0.9}$[CoO$_2$] and 
Na$_x$Ca$_y$[CoO$_2$], were reported to exhibit 
a small magnetic anomaly around 300~K,
\cite{NCO_6,NCO_7} 
probably indicating the existence of 
a spin state transition,
although the [CoO$_2$] plane was considered to be rigid
due to a network of edge-sharing CoO$_6$ octahedra.
Hence, the Co ions in the [CoO$_2$] planes are most likely to change
their spin state at 400 - 600~K.

The existence of the spin state transition suggests that 
the crystal-field splitting between $t_{2g}$ and $e_g$ levels 
is comparable to $\sim 400$~K. 
Thus, above $T_{\sf SS}^{\rm end}$,
charge carrier transport in the [CoO$_2$] plane 
is considered to be dominated 
by not only the direct hopping of the $t_{2g}$ electrons 
between the Co ions \cite{NCO_5}
but also the hybridization 
between Co-$e_g$ levels and  O-2$p$ levels, 
as in the case of the perovskite LaCoO$_3$.\cite{LCO_3}
Indeed, $\rho$ of LaCoO$_3$ decreased 
with increasing $T$ above 500~K,\cite{LCO_4} 
which is the temperature of the apparent spin state transition
from $LS$ to $IS$  
accompanied with an insulator-to-metal transition. 

If we employ the modified Heikes formula using the degeneracy of 
spin and orbital degrees of freedom of Co ions,\cite{Koshibae_1} 
the value of $S_{T\rightarrow\infty}$  
is calculated as 154~$\mu$VK$^{-1}$,
when the concentration of Co$^{4+}$ is 
equivalent to that of Co$^{3+}$ 
and both Co$^{3+}$ and Co$^{4+}$ are in the $LS$ state.
In order to explain the observed $S(T)$ curve 
above $T_{\sf SS}^{\rm end}$,
it is necessary to assume that  
Co$^{3+}$ is in the $LS$ state and Co$^{4+}$ in the $LS+IS$ state;
under this assumption, we obtain 
$S_{T\rightarrow\infty}$=293~$\mu$VK$^{-1}$,
although such inequivalent spin state is unlikely to exist
at elevated temperatures.

Our $\chi$ measurements using single crystal platelets of 
[Ca$_2$CoO$_3$]$_{0.62}$[CoO$_2$]
showed a clear thermal hysteresis with a width of $\sim$ 25~K 
for the spin state transition at $T_{\sf SS}^{\rm end}$.
\cite{jun_PRB2} 
The thermal hysteresis was also confirmed 
by a heat capacity measurement.\cite{Asahi_1}
These indicate that the spin state transition accompanies 
a structural change, 
as well as the case of LaCoO$_3$ around 100~K and 500~K
detected by neutron diffraction measurements.
\cite{LCO_5,LCO_6} 

Hence, in order to elucidate the mechanism of the spin state transition,
further $\mu^+$SR experiments on the related cobaltites,
such as Na$_{0.9}$[CoO$_2$] and 
Na$_x$Ca$_y$[CoO$_2$],
are necessary; 
in particular, a precise muonic Knight shift measurement
would provide a significant information on the change in $H_{\rm int}$ 
during the spin state transition.
In addition, we need an x-ray and/or neutron diffraction analysis for
[Ca$_2$CoO$_3$]$_{0.62}^{\rm RS}$[CoO$_2$]
as a function of temperature, 
to obtain the information on structural change, 
which would affect the magnitude and distribution of $H_{\rm int}$ 
above $T_{\sf SS}^{\rm end}$.
Furthermore, 
the photo-emission and x-ray absorption studies on 
[Ca$_2$CoO$_3$]$_{0.62}$[CoO$_2$]
at elevated temperatures 
are needed to determine the spin state
for understanding the transport properties 
above $T_{\sf SS}^{\rm end}$.

\section{\label{sec:Summary} Summary}
\begin{figure}
\vspace*{2\baselineskip}
\includegraphics[width=8cm]{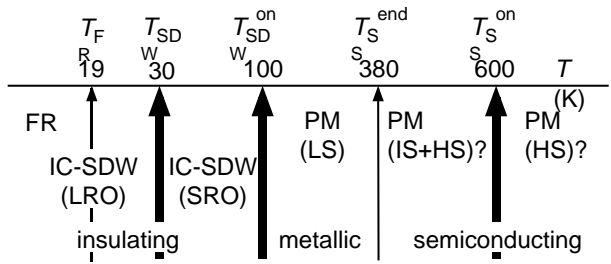} 
\caption{\label{fig:MagnTrans} 
  Successive magnetic transitions in [Ca$_2$CoO$_3$]$_{0.62}$[CoO$_2$].  
  The bold arrows indicate the transitions found by 
  the present and the previous $\mu^+$SR experiments,
  \cite{jun_PRB1,jun_PhysicaB1} 
  while the narrow arrows show those detected by 
  the previous susceptibility measurements.\cite{CCO_2,jun_PRB1}
  In Fig.~\ref{fig:MagnTrans}, 
  {\sf FR} means ferrimagnetic,
  {\sf PM} paramagnetic and 
  {\sf LS}, {\sf IS} and {\sf HS} 
  low-, intermediate- and high-spin state, respectively;
  and 
  {\sf IC-SDW} incommensurate spin density wave state, 
  {\sf LRO} and {\sf SRO} long-range and short-range order.
  The spin states above $T_{\sf SS}^{\rm end} \approx 380$~K 
  are not clear at present.  
}
\end{figure}
We investigated the magnetism of 
the misfit layered cobaltite,
[Ca$_2$CoO$_3$]$_{0.62}$[CoO$_2$],
by a positive muon spin rotation and relaxation experiment.
It is found that 
[Ca$_2$CoO$_3$]$_{0.62}$[CoO$_2$]
exhibits the successive magnetic transitions,
as summarized in Fig.~\ref{fig:MagnTrans}.  
An incommensurate ({\sf IC}) spin density wave ({\sf SDW}) 
is observed directly by ZF-$\mu^+$SR below about 30~K, 
and evidence for the onset of the {\sf IC-SDW} state is seen 
below $T_{\sf SDW}^{\rm on} \approx 100$~K, 
while the muon spin relaxation is characteristic of 
a paramagnet (PM) above $T_{\sf SDW}^{\rm on}$.  
Therefore, we conclude that 
the long-range {\sf IC-SDW} order completes 
below $\sim$ 30~K, 
whereas the short-range order appears below 100~K.
Also the {\sf IC-SDW} is found to
propagate in the [CoO$_2$] plane, 
with oscillating moments directed along the $c$-axis. 
Below $T_{\sf FR} \approx 19$~K, the {\sf IC-SDW} 
apparently coexists with ferrimagnetism ({\sf FR}).  

At 400 - 600~K, the spin state of Co ions 
changes; that is,
the populations of the low-, intermediate- and high-spin states 
are most likely to vary gradually with increasing temperature 
above 380~K (= $T_{\sf SS}^{\rm end}$).
Also, this transition is found to be sensitive to the Co valence,
which is controlled by doping and/or oxygen deficiency.

These two magnetic transitions detected by $\mu^+$SR 
are found to correlate closely with 
the transport properties of 
[Ca$_2$CoO$_3$]$_{0.62}$[CoO$_2$];
in particular, both a broad minimum 
at around 80~K and
a broad maximum between 
400 and 600~K
in the $\rho(T)$ curve.

\begin{acknowledgments} 
We thank Dr. S.R. Kreitzman, Dr. B. Hitti and Dr. D.J. Arseneau 
of TRIUMF for help with the $\mu^+$SR experiments.  
Also, we thank Mr. A. Izadi-Najafabadi and Mr. S.D. LaRoy 
of University of British Columbia for help with the experiments. 
We appreciate useful discussions with 
Dr. R. Asahi of Toyota Central R\&D Labs., Inc.,  
Prof. U. Mizutani, Prof. H. Ikuta and Prof. T. Takeuchi
of Nagoya University and 
Prof. K. Machida of Okayama University. 
This work was supported 
at Toyota CRDL by joint research and development with
International Center for Environmental Technology Transfer in 2002-2004,
commissioned by the Ministry of Economy Trade and Industry of Japan,
at UBC by the Canadian Institute for Advanced Research, 
the Natural Sciences and Engineering Research Council of Canada, 
and at TRIUMF by the National Research Council of Canada.  
\end{acknowledgments}


\end{document}